\documentclass[aps,prc,twocolumn,nofootinbib,superscriptaddress]{revtex4} 

\usepackage{graphicx,longtable,epsfig} %
\usepackage{dcolumn}
\usepackage{mathptmx}
\usepackage{amsmath}
\usepackage{epstopdf}
\usepackage[pdftex]{hyperref}
\usepackage{sidecap}
\sidecaptionvpos{figure}{c}
\usepackage[sort&compress]{natbib}

\usepackage{todonotes}

\begin{document}

\title{Enhanced low-energy $\gamma$-decay strength 
of $^{70}$Ni and its robustness within the shell model
}

\author{A.~C.~Larsen}
\email{a.c.larsen@fys.uio.no}
\affiliation{Department of Physics, University of Oslo, N-0316 Oslo, Norway}
\author{J.~E.~Midtb{\o}}
\email{j.e.midtbo@fys.uio.no}
\affiliation{Department of Physics, University of Oslo, N-0316 Oslo, Norway}

\author{M.~Guttormsen}
\affiliation{Department of Physics, University of Oslo, N-0316 Oslo, Norway}

\author{T.~Renstr{\o}m}
\affiliation{Department of Physics, University of Oslo, N-0316 Oslo, Norway}

\author{S.~N.~Liddick}
\affiliation{National Superconducting Cyclotron Laboratory, 
    Michigan State University, East Lansing, Michigan 48824, USA}
\affiliation{Department of Chemistry, Michigan State University, 
    East Lansing, Michigan 48824, USA}

\author{A.~Spyrou}
\affiliation{National Superconducting Cyclotron Laboratory, 
    Michigan State University, East Lansing, Michigan 48824, USA}
\affiliation{Department of Physics and Astronomy, 
    Michigan State University, East Lansing, Michigan 48824, USA}
\affiliation{Joint Institute for Nuclear Astrophysics, 
    Michigan State University, East Lansing, Michigan 48824, USA}

\author{S.~Karampagia}
\affiliation{National Superconducting Cyclotron Laboratory, 
    Michigan State University, East Lansing, Michigan 48824, USA}

\author{B.~A.~Brown}
\affiliation{National Superconducting Cyclotron Laboratory, 
    Michigan State University, East Lansing, Michigan 48824, USA}
\affiliation{Department of Physics and Astronomy, 
    Michigan State University, East Lansing, Michigan 48824, USA}
\author{O.~Achakovskiy}
\affiliation{Institute of Physics and Power Engineering,
249033 Obninsk, Russia}

\author{S.~Kamerdzhiev}
\affiliation{National Research Centre “Kurchatov Institute”, 123182 Moscow,  Russia}

\author{D.~L.~Bleuel}
\affiliation{Lawrence Livermore National Laboratory, Livermore, California 94551, USA}

\author{A.~Couture}
\affiliation{Los Alamos Neutron Science Center, 
    Los Alamos National Laboratory, Los Alamos, New Mexico 87545, USA}

\author{L.~Crespo~Campo}
\affiliation{Department of Physics, University of Oslo, N-0316 Oslo, Norway}

\author{B.~P.~Crider}
\affiliation{National Superconducting Cyclotron Laboratory, 
    Michigan State University, East Lansing, Michigan 48824, USA}
\affiliation{Department of Physics and Astronomy, Mississippi State University, Mississippi State, MS 39762, USA}

\author{A.~C.~Dombos}
\affiliation{National Superconducting Cyclotron Laboratory, 
    Michigan State University, East Lansing, Michigan 48824, USA}
\affiliation{Department of Physics and Astronomy, 
    Michigan State University, East Lansing, Michigan 48824, USA}
\affiliation{Joint Institute for Nuclear Astrophysics, 
    Michigan State University, East Lansing, Michigan 48824, USA}
    
\author{R.~Lewis}
\affiliation{National Superconducting Cyclotron Laboratory, 
    Michigan State University, East Lansing, Michigan 48824, USA}
\affiliation{Department of Chemistry, Michigan State University, 
    East Lansing, Michigan 48824, USA}
\author{S.~Mosby}
\affiliation{Los Alamos Neutron Science Center, 
    Los Alamos National Laboratory, Los Alamos, New Mexico 87545, USA}

\author{F.~Naqvi}
\affiliation{National Superconducting Cyclotron Laboratory, 
    Michigan State University, East Lansing, Michigan 48824, USA}
    
\author{G.~Perdikakis}
\affiliation{Department of Physics, Central Michigan University, Mount Pleasant, Michigan, 48859, USA}
\affiliation{National Superconducting Cyclotron Laboratory, 
    Michigan State University, East Lansing, Michigan 48824, USA}
\affiliation{Joint Institute for Nuclear Astrophysics, 
    Michigan State University, East Lansing, Michigan 48824, USA}
    
\author{C.~J.~Prokop}
\affiliation{National Superconducting Cyclotron Laboratory, 
    Michigan State University, East Lansing, Michigan 48824, USA}
    
\author{S.~J.~Quinn}
\affiliation{National Superconducting Cyclotron Laboratory, 
    Michigan State University, East Lansing, Michigan 48824, USA}
\affiliation{Department of Physics and Astronomy, 
    Michigan State University, East Lansing, Michigan 48824, USA}
\affiliation{Joint Institute for Nuclear Astrophysics, 
    Michigan State University, East Lansing, Michigan 48824, USA}

\author{S.~Siem}
\affiliation{Department of Physics, University of Oslo, N-0316 Oslo, Norway}

\date{\today}

\begin{abstract}
Neutron-capture reactions on very neutron-rich nuclei are essential for heavy-element nucleosynthesis through the rapid neutron-capture process, now shown to take place in neutron-star merger events. 
For these exotic nuclei, radiative neutron capture is extremely sensitive to their $\gamma$-emission probability at very low $\gamma$ energies. 
In this work, we present measurements of the $\gamma$-decay strength of $^{70}$Ni 
over the wide range $1.3 \leq E_{\gamma} \leq 8 $ MeV. 
A significant enhancement is found in the $\gamma$-decay strength  for transitions with $E_\gamma < 3$ MeV. 
At present, this is the most
neutron-rich nucleus displaying this feature, proving that this phenomenon is not restricted to stable nuclei. We have performed $E1$-strength calculations within the quasiparticle time-blocking approximation, which describe our data above $E_\gamma \simeq 5$ MeV very well. Moreover, large-scale shell-model calculations indicate an $M1$ nature of the
low-energy $\gamma$ strength. This turns out to be remarkably robust with respect to the choice of interaction, truncation  and model space, and we predict its presence in the whole isotopic chain, in particular the neutron-rich $^{72,74,76}\mathrm{Ni}$.

\end{abstract}

\maketitle

\section{Introduction} 
\label{sec:intro}
One of the most intriguing and long-standing scientific quests 
is the understanding of the fundamental building blocks in nature. 
Indeed, new paradigms have been established as new and improved measurements have been 
made available. A striking example is the standard model of particle physics~\cite{oerter2006}, proven to be extremely 
robust and predictive. On the nuclear scale, significant progress has been made as well, but a unified theory describing all facets of nuclear structure and dynamics for all nuclei is still lacking (see, \textit{e.g.}, Ref.~\cite{ring2004}). 

A particularly challenging task is to properly describe nuclear properties in the energy
regime where the average spacing, $D$, between the available quantum states is still larger than
the width $\Gamma$ of the state, but so small that conventional spectroscopy is impractical
or nearly impossible. This region, generally known as the \textit{quasi-continuum}, is of
particular interest for studying nuclear dynamics such as breaking of nucleon-Cooper
pairs~\cite{melby1999}, as well as $\gamma$-decay resonances (see, \textit{e.g.}, Refs.~\cite{krticka2004,schiller2006,heyde2010,savran2013} and references therein). 

In addition to the pure nuclear-structure motivation, 
the quasi-continuum is of vital importance to properly describe and understand
the creation of elements heavier than iron~\cite{go98,ar07}, 
which has been identified as one of the ``\textit{Eleven Science Questions for the New Century}''~\cite{questions2003}.
A clear signature of the \textit{rapid neutron-capture process} ($r$-process) has finally been observed: Gravitational waves from a neutron-star merger event were observed with the Advanced LIGO and Advanced Virgo detectors~\cite{LIGO2017}, and electromagnetic counterparts show that the $r$-process has indeed taken place in this event~\cite{drout2017}.
Our detailed understanding of the $r$-process is, however, still largely hampered by 
the lack of crucial nuclear-physics input, such as masses, $\beta$-decay probabilities, and 
radiative neutron-capture, ($n,\gamma$), rates~\cite{ar07}. 

Two of the needed nuclear properties for understanding nuclear dynamics
in the quasi-continuum as well as 
calculating astrophysical $(n,\gamma)$ reaction rates are the nuclear level density (NLD)
and the $\gamma$-strength function ($\gamma$SF). The former is simply the average number of quantum
levels per energy bin as a function of excitation energy, 
while the latter is a measure of the \textit{average}, reduced 
$\gamma$-decay probability. The $\gamma$SF is dominated 
by the $E1$ Giant Dipole Resonance (GDR, \textit{e.g}, Refs.~\cite{dietrich1988,harakeh2000}).
 \begin{figure*}[ht]
 \begin{center}
 \includegraphics[clip,width=2.15\columnwidth]{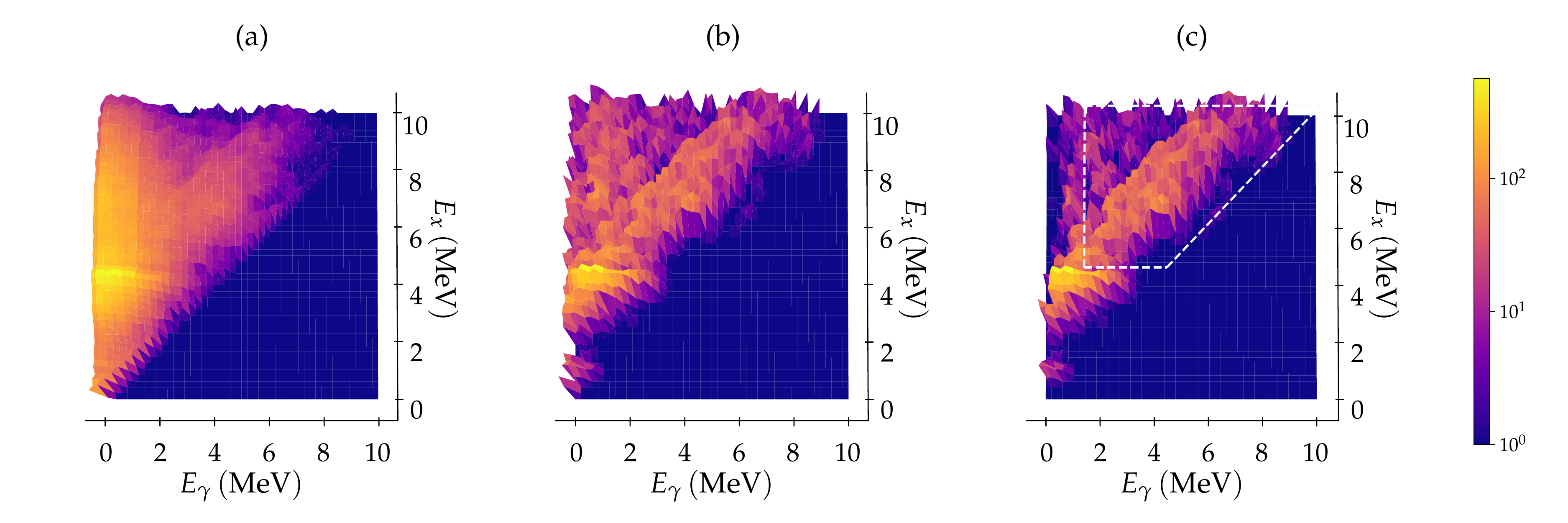}
 \caption {(Color online) (a) $^{70}$Ni raw $\gamma$ spectra versus $E_x$; 
(b) unfolded $\gamma$ spectra corrected for the SuN response functions for both $E_\gamma$ and $E_x$; 
(c) distribution of primary $\gamma$ rays for each $E_x$ bin. The dashed lines show the region
 used for the analysis. The pixels are 200 keV wide.
 }
 \label{fig:matrices}
 \end{center}
 \end{figure*}

In recent years, an unexpected enhancement in the $\gamma$SF at low $\gamma$ energies 
($E_\gamma <3-4$ MeV) has been observed in many $fp$-shell
and medium-mass nuclei (\textit{e.g.}, Refs.~\cite{Fe_Alex,larsen2013,Mo_RSF,wiedeking95Mo,renstrom2016}), with $^{138-140}$La~\cite{kheswa2015,kheswa2017} and $^{151,153}$Sm~\cite{simon2016} being the heaviest so far. 
The multipolarity of this low-energy enhancement, referred to as the \textit{upbend} in the following, 
has been experimentally verified to be of dipole type~\cite{larsen2013,simon2016,larsen2017}.
However, theoretical attempts to describe the upbend differ on the underlying mechanism
and electromagnetic character. The authors of Ref.~\cite{litvinova2013} find 
an enhancement in the theoretical $E1$ strength, while shell-model approaches~\cite{schwengner2013,brown2014,schwengner2017}
demonstrate a large low-energy $M1$ enhancement at high excitation energies.
Shell model calculations including both $E1$ and $M1$ components confirm the $M1$ upbend, but also predict the $E1$ $\gamma$SF to be constant for low energies \cite{Sieja2017}.
A recent experiment on $^{56}$Fe, although hampered by limited statistics, indicates that the low-energy enhancement could be a mix of both components, with a small magnetic bias between 1.5 and 2 MeV \cite{Jones2018}.

Turning to the $r$-process, the presence of an upbend  
could increase the astrophysical ($n,\gamma$) reaction rates up to $\sim 2$ orders of magnitude
for very neutron-rich nuclei~\cite{larsen_goriely}.
Prompt neutron-star merger ejecta correspond to a cold and neutron-rich $r$-process where an $(n,\gamma)-(\gamma,n)$ equilibrium will never be established~\cite{ar07,just2015}. Hence, $(n,\gamma)$ rates will have a significant impact on the $r$-process reaction flow and final abundance distribution. It is therefore crucial to understand the nature of this upbend and search for its presence in nuclei far from stability.

In this article, we present NLD and $\gamma$SF measurements
of the neutron-rich nucleus $^{70}$Ni, using the newly developed $\beta$-Oslo method~\cite{spyrou2014,liddick2016}. 
Furthermore, we have  calculated the $E1$ strength within the quasiparticle time-blocking approximation (QTBA), and performed large-scale shell-model (SM) calculations for a wide range of effective interactions and model spaces,
exploring the $M1$ strength within this framework. 
We find that the upbend is indeed explained by the shell-model calculations, and we predict its presence in the whole isotopic chain, in particular the neutron-rich $^{72,74,76}\mathrm{Ni}$.

\section{Experimental details and data analysis} 
\label{sec:exp}
The experiment has already been described in Refs.~\cite{liddick2016,Spyrou2016,spyrou2017}; a brief summary is given in the following. The experiment was conducted at the 
National Superconducting Cyclotron Laboratory, Michigan State University,
where $^{70}$Co fragments were produced from 
a primary beam of $^{86}$Kr with energy 140 MeV/A impinging on a 
$\approx 400$ mg/cm$^2$ Be target and
selected with the A1900 fragment separator~\cite{morrissey2003}. 
The fragments were implanted in a double-sided silicon strip detector (DSSD) 
of 1-mm thickness mounted in the center of the Summing NaI (SuN) total absorption spectrometer~\cite{simon2013}. SuN is a large-volume barrel consisting of eight optically isolated 
segments, providing information on the individual $\gamma$ rays, while the sum of all detected 
$\gamma$ rays gives the initial excitation energy of the daughter nucleus.
Coincidences between $\beta^-$ particles
and the fragment were determined by the DSSD using the implantation and $\beta$-decay pixel positions in the DSSD and absolute times of the signals. 
The $\gamma$ rays measured with SuN were gated on 
the implantation-$\beta$-decay events to obtain the $\gamma$-ray spectra of the daughter nuclei.  
The summing efficiency of SuN varies with $\gamma$ multiplicity and initial excitation energy, and is, 
on average, $\approx 25-30$\%. 

The individual $\gamma$-ray spectra versus the summed $\gamma$-ray energies ({\it i.e.}, initial excitation energy $E_x$)
of $^{70}$Ni are shown in Fig.~\ref{fig:matrices}a; the total number of counts are about 72,000. 
The $\gamma$ spectra were unfolded along the $E_\gamma$ axis with the technique 
described in Ref.~\cite{guttormsen1996} using SuN response functions generated with GEANT4~\cite{geant4,geant4_2} simulations of the
full setup. 
Furthermore, due to the possibility of incomplete summing and a high-energy tail induced by electrons from the $\beta$-decay ($Q_\beta = 12.3$ MeV), we have also developed an unfolding technique for the summed $\gamma$-rays. 
This technique is based on the one in Ref.~\cite{guttormsen1996} and will be presented thoroughly in a forthcoming article~\cite{guttormsen_systbeta}. The resulting  unfolded
matrix is shown in Fig.~\ref{fig:matrices}b.

After unfolding, the distribution of  primary $\gamma$ rays for each 200-keV $E_x$ bin 
was extracted by an iterative subtraction technique described in detail in Refs.~\cite{Gut87,larsen2011}. 
The basic principle behind this technique is that, for a given excitation-energy bin $E_j$, 
the distribution of the first-emitted $\gamma$ rays ({\it i.e.}~branchings) is determined by subtracting the 
$\gamma$-ray spectra from the lower excitation-energy bins $E_{i<j}$. This is true if, for a given $E_x$ bin,
approximately the same spin distribution is populated directly from the $\beta$-decay and 
by the $\gamma$ decay into this bin from above. 

Previous experiments have shown that 
the ground-state spin/parity of $^{70}$Co is $(6^-,7^-)$~\cite{mueller2000,sawicka2003}. 
Assuming a spin/parity of $6^-$, the $\beta$-decay will mainly
populate levels with spin/parity $5^-, 6^-, 7^-$ in the initial $E_x$ bins through Gamow-Teller transitions.
With one dipole $\gamma$ transition either of electric or magnetic type, 
the spins populated in the underlying bins are $J=4-8$ (both parities). 
On the other hand, if the $^{70}$Co ground state has spin/parity $7^-$, the initial levels of $^{70}$Ni will have  spin/parity $6^-, 7^-, 8^-$, and the final levels following one dipole transition will be $J=5-9$. 
Further, although the timing requirements in the data analysis strongly favors population in $^{70}$Ni from the short-lived $\approx100$-ms ($6^-,7^-$) level in $^{70}$Co, a small contribution from the longer-lived $\approx 500$-ms ($3^+$) level could be present. 
We also note that in a recent study of the decay chain $^{70}$Fe$\rightarrow$$^{70}$Co$\rightarrow$$^{70}$Ni by Morales \textit{et al.}~\cite{morales2017}, it is suggested that the spin/parity of the longer-lived level could be ($1^+,2^+$). 

 \begin{figure}[tb]
 \begin{center}
 \includegraphics[clip,width=1.\columnwidth]{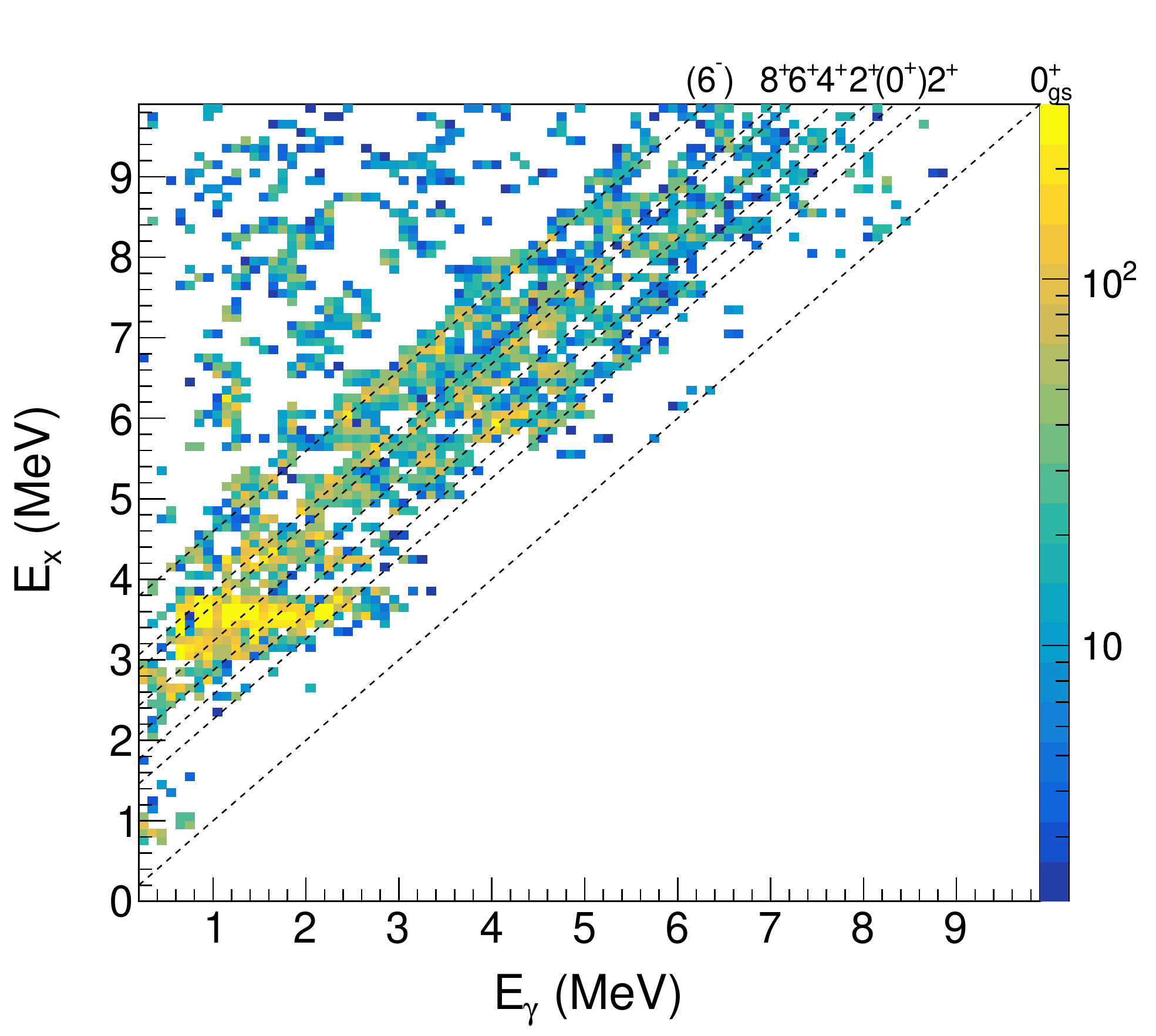}
 \caption {(Color online) Matrix of primary $\gamma$ rays as a function of excitation energy. The dashed lines show the direct decay to the low-energy levels marked with their spin/parity assignment (see text). The bin width is 100 keV both on the $\gamma$-ray and $E_x$ axis.}
 \label{fig:fgmatrix}
 \end{center}
 \end{figure}
In Fig.~\ref{fig:fgmatrix}, we display the matrix of primary $\gamma$ rays for each excitation-energy bin, where the decay to some of the low-energy levels are indicated with the dashed lines. 
It is obvious that there is no direct decay to the ground state or to the ($0^+$) state at 1567 keV~\cite{prokop2015}, as can be expected from an initial spin population of $J>1$. 
Thus, it is doubtful that the long-lived level in $^{70}$Co is ($1^+$) as indicated as a possibility in Ref.~\cite{morales2017}.
However, our data are fully consistent with both the suggested ($2^+$)~\cite{morales2017} and the ($3^+$)~\cite{mueller2000,prokop2015} assignments. 
We will in the following use ($2^+,3^+$) for the spin/parity assignment of this level.
Further, we observe some direct $\gamma$ decay to the second $2^+$ level at 1866 keV, as well as a much weaker direct $\gamma$ decay to the first $2^+$ level at 1259 keV. 
This indicates that even though the $\beta$ decay from the $^{70}$Co ($6^-,7^-$) level is the dominant component, there is also a weaker contribution from the ($2^+,3^+$) long-lived state in our data. 
By inspection of the decay curve, we find that the ($2^+,3^+$) contribution is indeed small, of the order of $5-10$\%.

Moreover, we observe direct $\gamma$-decay to the $4^+$ level at $E_x = 2229$ keV, which could be reached through $E1$ transitions from $5^-$ levels, or by $M1$ transitions from $3^+$ and $4^+$ levels populated from the long-lived ($2^+,3^+$) state in $^{70}$Co.
As the strongest decay to the $4^+$ level is seen at $E_x \approx 5.4$ and 6.0 MeV, one would expect $E1$ dominance at such high excitation energies, which is further supported by our calculations presented in Sec.~\ref{sec:theory}.
Hence, we find that $^{70}$Co most likely has spin/parity $6^-$ in its ground state, although the $7^-$ assignment cannot be completely ruled out.

 \begin{figure}[tb]
 \begin{center}
 \includegraphics[clip,width=1.\columnwidth]{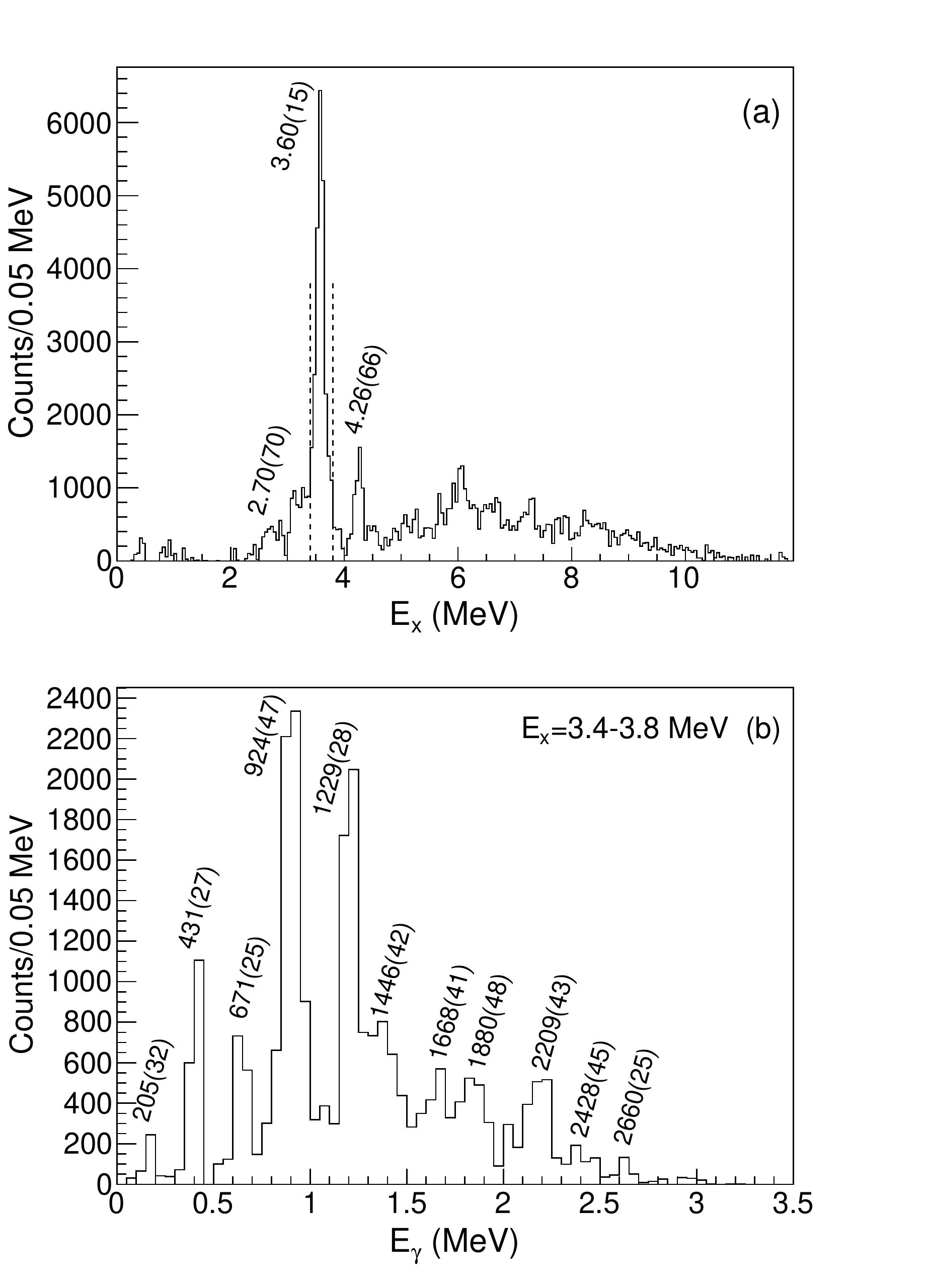}
 \caption {Projections of the unfolded $E_x - E_\gamma$ matrix onto (a) the $E_x$ axis, and (b) the $E_\gamma$ axis for a gate on $E_x = 3.4-3.8$ MeV (dashed lines in the upper panel). 
 Transitions are labeled with their $\gamma$-ray energies in keV.}
 \label{fig:unf_projections}
 \end{center}
 \end{figure}
In Fig.~\ref{fig:unf_projections}a we show the projection of the unfolded $\gamma$-ray matrix onto the excitation-energy axis. 
This spectrum represents the distribution of level population in $^{70}$Ni through $\beta$ decay of $^{70}$Co, and effectively demonstrates that there is no direct population of levels below $E_x \approx 2.5$ MeV. 
This proves that there is no direct feeding from the ($2^+,3^+$) level to the low-lying levels. 
Further, in Fig.~\ref{fig:unf_projections}b, we have projected the unfolded $\gamma$-ray matrix onto the $\gamma$-energy axis, showing all transitions in the $\gamma$-decay cascades for an excitation-energy gate of $E_x = 3.4-3.8$ MeV. 
This gate includes the cascades from the strongly populated ($6^-$) level at 3592 keV stemming from the ($6^-,7^-$) $^{70}$Co ground state, as well as a contribution from the 3510-keV level populated via the ($2^+,3^+$) long-lived state. 
We clearly see strong lines that can be identified (within their uncertainties) to known decay cascades of the ($6^-$) level~\cite{prokop2015}, but in addition we see weaker transitions of higher $\gamma$-ray energies, which are likely originating from the level at 3510 keV populated from the ($2^+,3^+$) level.

For the following analysis and comparison with the theoretical calculations, we would like to stress that there is no major change in the conclusions drawn if the ($7^-$) spin/parity assignment turns out to be the correct one.
Further, the contribution from the ($2^+,3^+$) level in $^{70}$Co is quite small compared to the ($6^-,7^-$) one, as demonstrated by the dominance of decay to higher-spin levels in Figs.~\ref{fig:fgmatrix} and~\ref{fig:unf_projections}b.
However, the very strong direct population of the $(6^-)$ level at 3592 keV and the non-population of lower-lying levels could cause problems in the subtraction technique to obtain the primary $\gamma$-ray spectra.
Further, considering that the initial levels are dominantly populated from the $(6^-,7^-)$ level we will expect, on average, to subtract somewhat too much of $\gamma$ rays below $\approx 3$ MeV, as the 
underlying $E_x$ bins will contain $\gamma$ rays from a broader spin range than what is 
populated directly through the $\beta$-decay. Indeed, this is also what we observe in Figs.~\ref{fig:matrices}c and~\ref{fig:fgmatrix}: the higher-energy primary $\gamma$ rays are not much affected as they are 
dominantly primary transitions, but
we clearly see that there is a region for $E_x \approx 7-9$ MeV where there are few
low-energy $\gamma$ rays. 
This will lead to a poor estimate of the NLD at high excitation energies,
but will not hamper the extraction of the $\gamma$SF.

 \begin{figure*}[h!bt]
 \begin{center}
 \includegraphics[clip,width=2.\columnwidth]{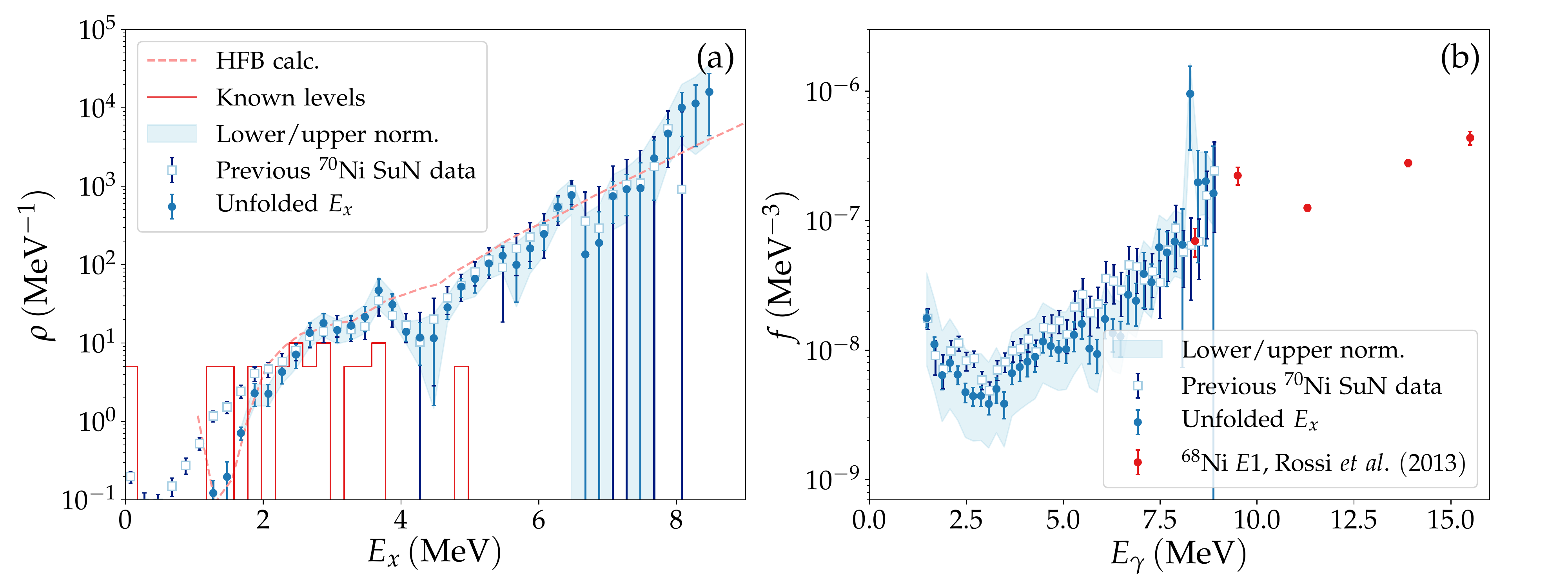}
 \caption {(Color online) (a) Extracted NLD for $^{70}$Ni from the analysis in Ref.~\cite{liddick2016} (open squares) and the present $E_x$-unfolded SuN data (blue points) with upper/lower limits 
   (blue shaded area) and the HFB+c calculations used for normalization (dashed line). 
(b) Extracted $\gamma$SF for $^{70}$Ni. The data of Rossi \textit{et al.}~\cite{rossi2013} on $^{68}$Ni (red points) 
    are used for normalization.
 }
 \label{fig:nld_gsf}
 \end{center}
 \end{figure*}

\section{Results}
\label{sec:results}
Having the distributions of primary $\gamma$ spectra on hand for each excitation-energy bin, 
we extracted the NLD and $\gamma$-transmission 
coefficient for $^{70}$Ni using the least $\chi^2$ method described in Ref.~\cite{Schiller00}.
The main principle of this method is to fit all data points in the selected region of the two-dimensional 
landscape of primary $\gamma$ rays with two functions; {\it i.e.}, the matrix of primary $\gamma$ rays 
$P(E_\gamma,E_x)$, normalized for each $E_x$ so that $\sum_{E_\gamma=0}^{E_x}P(E_\gamma,E_x)=1$,
can be described with the product $\rho(E_f)\cdot\mathcal{T}(E_\gamma)$ for the final excitation energy
$E_f = E_x-E_\gamma$. Here, $\rho(E_f)$ is the NLD and $\mathcal{T}$ is the $\gamma$ transmission coefficient; the $\gamma$SF for dipole strength, $f(E_\gamma)$, is derived from $\mathcal{T}$ through $f(E_{\gamma}) = {\mathcal T}(E_{\gamma})/2\pi E_{\gamma}^{3}$. 
Note that there are many more data points in the selected region than fit parameters; the 
simultaneous fit is thus providing a unique solution of the functional form of the NLD and $\gamma$SF. 

The extracted NLD and $\gamma$SF functions are normalized as described in Ref.~\cite{liddick2016}; for the NLD, 
we make use of known, discrete levels of $^{70}$Ni taken from Refs.~\cite{ENSDF,chiara2015,prokop2015}. 
and Hartree-Fock-Bogoliubov
plus combinatorial (HFB-c) calculations taken from Ref.~\cite{goriely2008} using 
 an $E_x$ shift 
$\delta = -0.6, -0.8, -1.0$ MeV. With these shifts we reproduce the appearance of the first negative-parity level within $\approx 300$ keV. The normalized 
NLD is displayed in the left panel of Fig.~\ref{fig:nld_gsf}.  
The $\gamma$SF is normalized to the recently measured $E1$ strength above the neutron threshold of 
$^{68}$Ni by Rossi~\textit{et al.}~\cite{rossi2013} and shown in the right panel of Fig.~\ref{fig:nld_gsf}. 

We observe that the present NLD displays a steeper slope at $E_x \approx 1.5-2$ MeV than previously due to the unfolding of the $E_x$ axis, thus achieving an excellent agreement with the discrete levels for $E_x \approx 2-3$ MeV. Also, the $\gamma$SF clearly displays an upbend consistent with the findings
in iron isotopes~\cite{Fe_Alex,larsen2013}
and $^{60,64,65,69}$Ni~\cite{Ni_Alex,crespo2016,crespo2017,spyrou2017}. 
This gives support to the hypothesis that the upbend is a
general feature, and is not restricted to (near-)stable nuclei. 
Moreover, although rather scarce statistics at the very highest $\gamma$ energies, our data indicate an increase in strength in the $E_\gamma \approx 8-9$ MeV region. 
This feature is consistent with the pygmy dipole 
strength found in $^{68}$Ni~\cite{rossi2013,wieland2009}. 
Hence, our data prove the existence of the upbend in $^{70}$Ni, and give a hint to the presence of a pygmy dipole resonance.

\section{Theoretical calculations and comparison with data}
\label{sec:theory}
The high-energy part of the $\gamma$SF
($E_\gamma > 4$ MeV) is expected to be dominated by the $E1$ tail of the GDR. 
To describe the GDR part, we have performed $E1$-strength calculations based on the self-consistent extended version of the theory of finite Fermi systems (ETFFS) within the quasiparticle time-blocking approximation (QTBA)~\cite{kamerdziev2004,tselyaev2007,achakovskiy2016,achakovskiy2015} using the BSk17 Skyrme force~\cite{goriely2004}. 
The advantage of this approach is that it includes self-consistently the quasiparticle random phase approximation (one-particle-one-hole excitations on the ground state), phonon-coupling effects, and a discretized form of the single-particle continuum spectrum. The Skyrme force is used to calculate the mean field, effective nucleon-nucleon interaction and phonon properties~\cite{achakovskiy2016}. We emphasize that phonon-coupling effects are crucial to obtain good agreement with data.\cite{achakovskiy2016,achakovskiy2015}

\begin{figure*}[tb]
\begin{center}
  \includegraphics[clip,width=1.7\columnwidth]{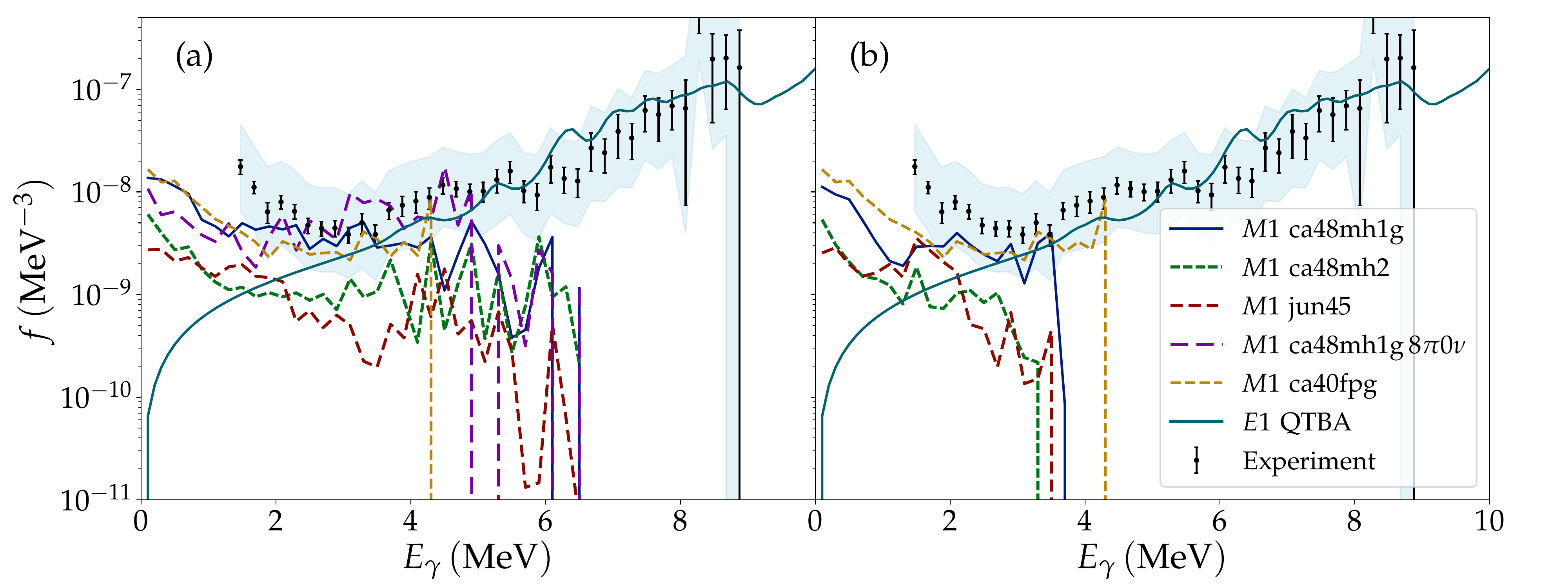}
\caption{(Color online) Calculated $\gamma$SFs within the shell model (for $4.0 \leq E_x \leq 6.5$ MeV) and the QTBA (for all $E_x$) compared with data. The blue shaded band indicates the total experimental uncertainty. In panel (a) all SM $M1$ transitions in the extraction region are included, while in (b) only transitions originating from levels with $J^\pi=5^-, 6^-$ or $7^-$ are shown (see text).}
\label{fig:gSF}
\end{center}
\end{figure*}

The resulting 
$E1$ strength is shown in Fig.~\ref{fig:gSF}. 
The agreement with the present data for $E_\gamma \approx 5-9$ MeV is excellent, within the experimental error bars. 
As the QTBA calculation is built on the ground state, this indicates that the average $E1$ strength between excited levels of $J_{\mathrm{initial}} = 5^-, 6^-, 7^-$ and $J_{\mathrm{final}} = 4^+-8^+$ is very similar to that of the $1^-$ levels decaying to the $0^+$ ground state, in accordance with the Brink-Axel hypothesis~\cite{brink1955,axel1962,guttormsen2016}.

To investigate the $M1$ radiation of $^{70}$Ni theoretically, we employ shell model calculations using the codes {\scshape KSHELL} \cite{Shimizu2013} and {\scshape NuShellX@MSU} \cite{Brown2014a}. To probe the robustness of the results, we use several different effective interactions. For the {\scshape KSHELL} calculations we use {\scshape jun45} \cite{Honma2009}, which contains the orbitals $\pi(p_{3/2}p_{1/2}f_{5/2}g_{9/2})$, $\nu(p_{3/2}p_{1/2}f_{5/2}g_{9/2})$; and two interactions called {\scshape ca48mh1} and {\scshape ca48mh2}, which include the $\pi f_{7/2}$ but exclude $\pi g_{9/2}$. The {\scshape ca48mh1} interaction is solely based on many-body perturbation theory (MBPT), {\it i.e.},~the two-body matrix elements (TBMEs) are not tuned to experimental data \cite{Hjorth-Jensen1995}. This is the same interaction that was used in a recent study of Fe isotopes \cite{schwengner2017}. Further, {\scshape ca48mh2} is derived from {\scshape ca48mh1} by replacing the neutron-neutron TBMEs with those of {\scshape jj44pna} \cite{Lisetskiy}, and modifying the diagonal $\pi f_{7/2}$ matrix elements based on the experimental spectrum of $^{54}$Fe. 

For the {\scshape NuShellX@MSU} calculations we use an interaction based on the {\scshape gxpf1a} $pf$ shell interaction \cite{Honma2002}, extended by MBPT-generated TBMEs to encompass the full $fpg$ model space for both protons and neutrons. This interaction was also used to predict $^{70}$Co $\beta$-decay intensities for the present experiment \cite{Spyrou2016}. In the following we will refer to this interaction as {\scshape ca40fpg}.

With the {\scshape ca48mh} model space, the full $M$-scheme basis size of $^{70}$Ni is $1.2 \times 10^9$ for each parity, and for {\scshape ca40fpg} it is even larger. For calculations in the {\scshape ca48mh} model space, we therefore restrict the maximum number of excited protons from the $f_{7/2}$ orbital to 2, as has been done in previous studies \cite{brown2014,schwengner2013,schwengner2017}, but with no truncations on neutrons. This reduces the basis size to $2 \times 10^7$. In the {\scshape ca40fpg} calculations we restrict the model space to the configurations $\pi \left( f_{7/2}^{8-t_p}(f_{5/2}pg)^{t_p} \right)$, $\nu\left( (fp)^{20-t_n} g_{9/2}^{2+t_n} g_{7/2}^0 \right)$ for $t_p, t_n = 0, 1$. For {\scshape jun45}, no truncation is applied. 
\begin{table}[bt]
\centering
   \begin{tabular}{l l l l l l}
        \\
                                               & Exp.    & {\scshape ca48mh1g} & {\scshape ca48mh2} & {\scshape jun45} & {\scshape ca40fpg} \\
       \hline
       \hline
       $B(E2; 2^+_1 \to 0^+_1)$ & 172(28) & 154.8          & 161.4         & 15.6  &  35.2 \\
       $B(E2; 6^+_1 \to 4^+_1)$ & 43(1)   & 120.0          & 230.7         & 5.7   &  24.5\\
       $B(E2; 8^+_1 \to 6^+_1)$ & 19(4)   & 21.7           & 139.7         & 2.2   &  9.5 \\
   \hline
   \end{tabular}
   \caption{Comparisons of yrast $B(E2)$ strengths (in units of e$^2$ fm$^4$) between experiment \cite{Perru} and SM calculations.}
   \label{table:Bstrengths}
\end{table}

\begin{figure*}[tb]
    \centering
    \includegraphics[width=1.9\columnwidth]{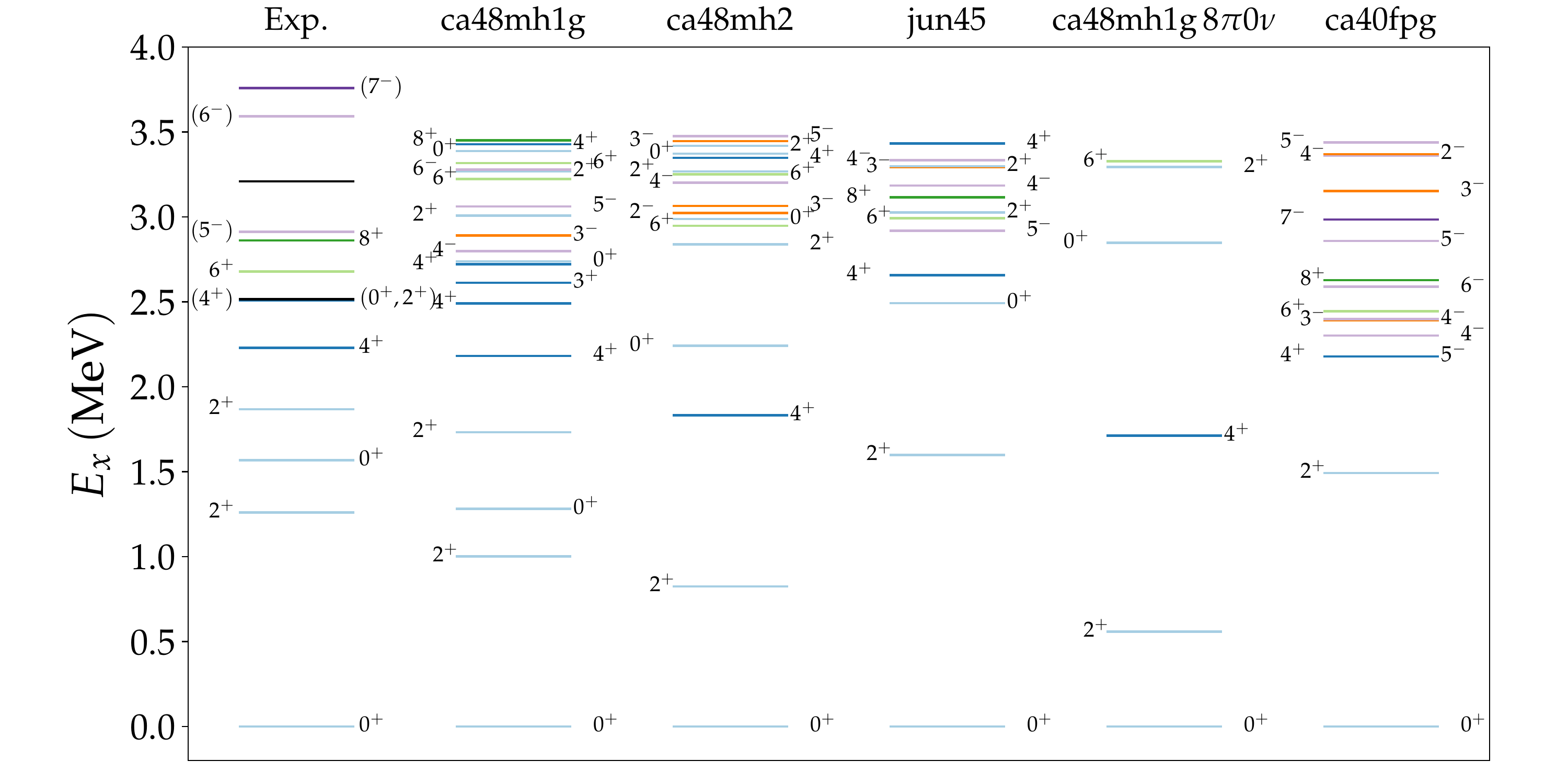}
    \caption{(Color online) Comparison of the experimental low-energy structure of $^{70}$Ni to calculations with {\tt ca48mh1g} and {\tt jun45}. The experimental data is from \cite{chiara2015} except the $0^+_2$ state which is revised according to \cite{prokop2015}.}
    \label{fig:low_lying_structure}
\end{figure*}

$^{70}$Ni exhibits a complex low-energy structure. The second excited state is $J^\pi = 0^+$ at $E(0^+_2)=1567$ keV \cite{prokop2015}, and calculations indicate it has a very different structure from the ground state \cite{chiara2015}. For the {\scshape ca48mh1} interaction we find good agreement with experiment by increasing the single-particle energy of the $\nu g_{9/2}$ orbital to 1.7 MeV. We will refer to the interaction with this modification as {\scshape ca48mh1g}. This interaction reproduces the low-lying spectrum to within a few hundred keV, 
including features such as the excited $0^+_2$ state and the onset of negative-parity states at $E_x \sim 3$ MeV. As shown in Table \ref{table:Bstrengths}, $B(E2)$ transitions strengths of the yrast band are also excellently reproduced, with the exception of the $B(E2; 6^+_1 \to 4^+_1)$ which is a factor of 3 too high. 
The drastic reduction in the experimental $B(E2)$ strength of the yrast band from $(2^+\to 0^+)$ to $(8^+\to 6^+)$ is discussed in \cite{Perru}, and is attributed there to core polarization by the tensor interaction that come into play for the lowest-lying states. 
The fact that we reproduce this transitional behaviour of the $B(E2)$ value
supports the applicability of the {\scshape ca48mh1g} interaction to this nucleus, at least for the low-lying levels. Predicted level schemes of the various interactions are shown in Fig.~\ref{fig:low_lying_structure}.

The {\scshape ca48mh2} interaction is complementary to {\scshape ca48mh1g} in that it overestimates the $0^+_2$ energy. It correctly predicts the $B(E2;2^+_1 \to 0^+_1)$ value, but fails to catch the transitional behaviour along the yrast line. The {\scshape jun45} calculation systematically overestimates level energies, and underestimates the $B(E2)$ strengths by an order of magnitude. Both calculations correctly reproduce the onset of negative parity states. The {\scshape ca40fpg} calculation does not reproduce the $0^+_2$ band, but this is to be expected due to the significant truncations applied to the model space. 

The fact that we have such a diverse ensemble of interactions that capture different features of $^{70}$Ni enables us to study the robustness of shell model calculations up to the quasi-continuum. We also probe the effects of model space truncations, using 0, 1 and 2 proton excitations from $f_{7/2}$ as well as varying neutron truncations. 

For each of the interactions, we calculate all states with $J\in[0,8]$ ($J\in[0,14]$), in the case of {\scshape NuShellX} ({\scshape KSHELL}) for both parities up to $S_n = 7.3$ MeV or above, and $B(M1)$ strengths of all allowed transitions between states. For the calculation of $B(M1)$ values, effective $g_s$ factors of $g_s = 0.9 g_s^\mathrm{free}$ have been used. We note that the recommended quenching factor for {\scshape JUN45} is $g_s = 0.9 g_s^\mathrm{free}$, because the core closure goes between spin-orbit partners ($f_{7/2}-f_{5/2}$). One could therefore argue that a somewhat larger quenching should be applied also for the $^{48}$Ca core interactions. This would serve to reduce the $M1$ strength function somewhat.

We extract the $\gamma$SF using the relation 
\begin{equation*}
   f_{M1}(E_\gamma, E_i, J_i, \pi_i) = a \langle B(M1)\rangle(E_\gamma, E_i, J_i, \pi_i) \rho(E_i, J_i, \pi_i), 
\end{equation*}
where $a=11.5473\times 10^{-9}\,\mu_N^{-2} \, \mathrm{MeV}^{-2}$, and $\rho(E_i, J_i, \pi_i)$ and $\langle B(M1) \rangle$ is the partial level density and the average reduced transition strength, respectively, of states with the given excitation energy, spin and parity \cite{bartholomew1972}. By the generalized Brink-Axel hypothesis, $f_{XL}(E_\gamma, E_i, J_i, \pi_i) \approx f_{XL}(E_\gamma)$. Hence we obtain $f_{M1}(E_\gamma)$ by averaging over $E_i, J$ and $\pi$. Only $E_i, J, \pi$ pixels where $f_{M1}$ is non-zero are included in the average.

\begin{figure}[tb]
\includegraphics[clip,width=1\columnwidth]{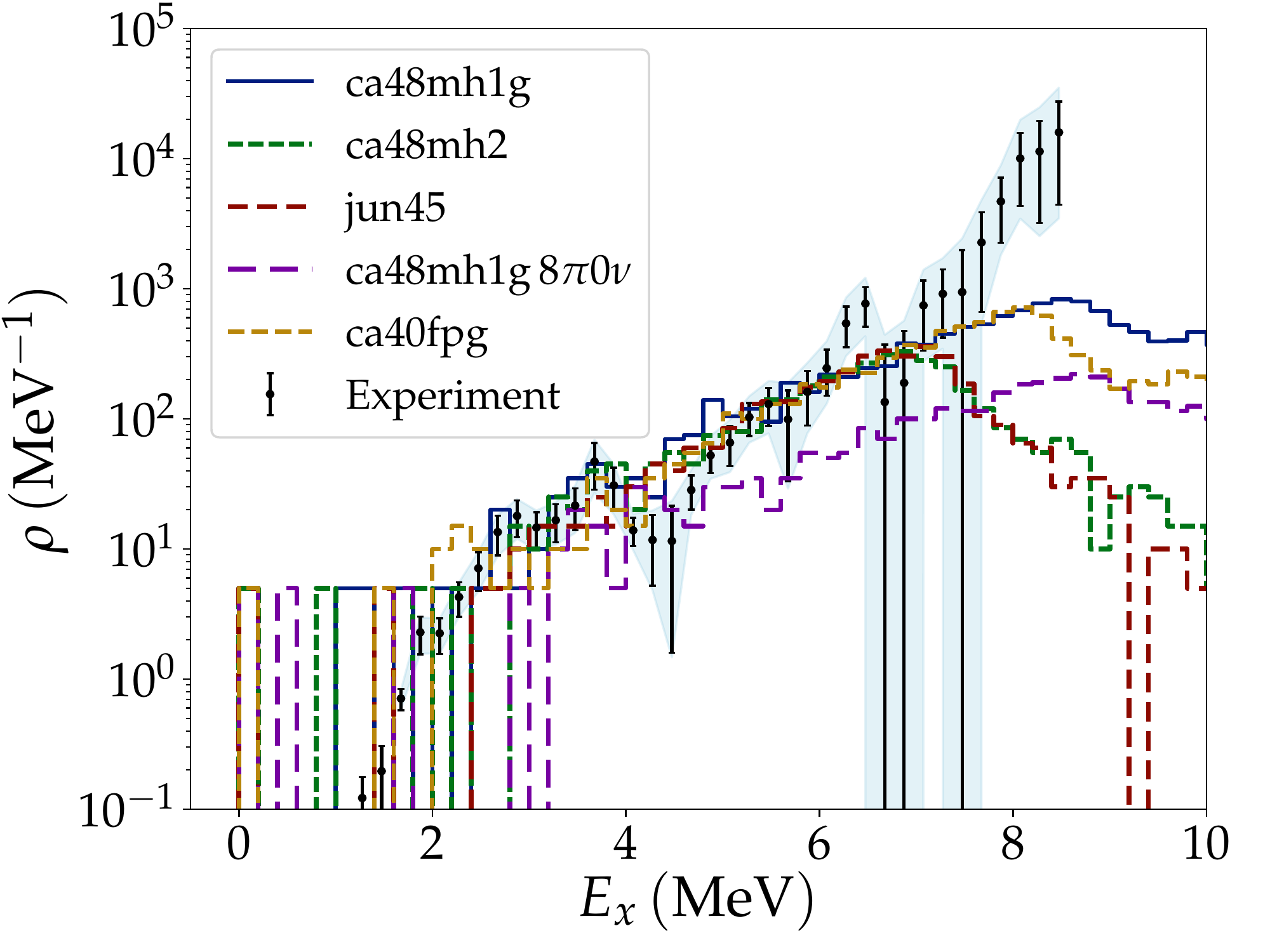}
\caption{(Color online) Calculated shell-model level densities compared to the $^{70}$Ni data. The grey band indicates the total experimental uncertainty, systematic and statistical.}
\label{fig:rho}
\end{figure}

We find that all SM calculations excellently match the experimental NLD up to $E_x \sim 6$ MeV, where they start to fall off because of the limited number of calculated states -- see Fig.~\ref{fig:rho}.

\begin{figure*}[tb]
\begin{center}
  \includegraphics[clip,width=1.7\columnwidth]{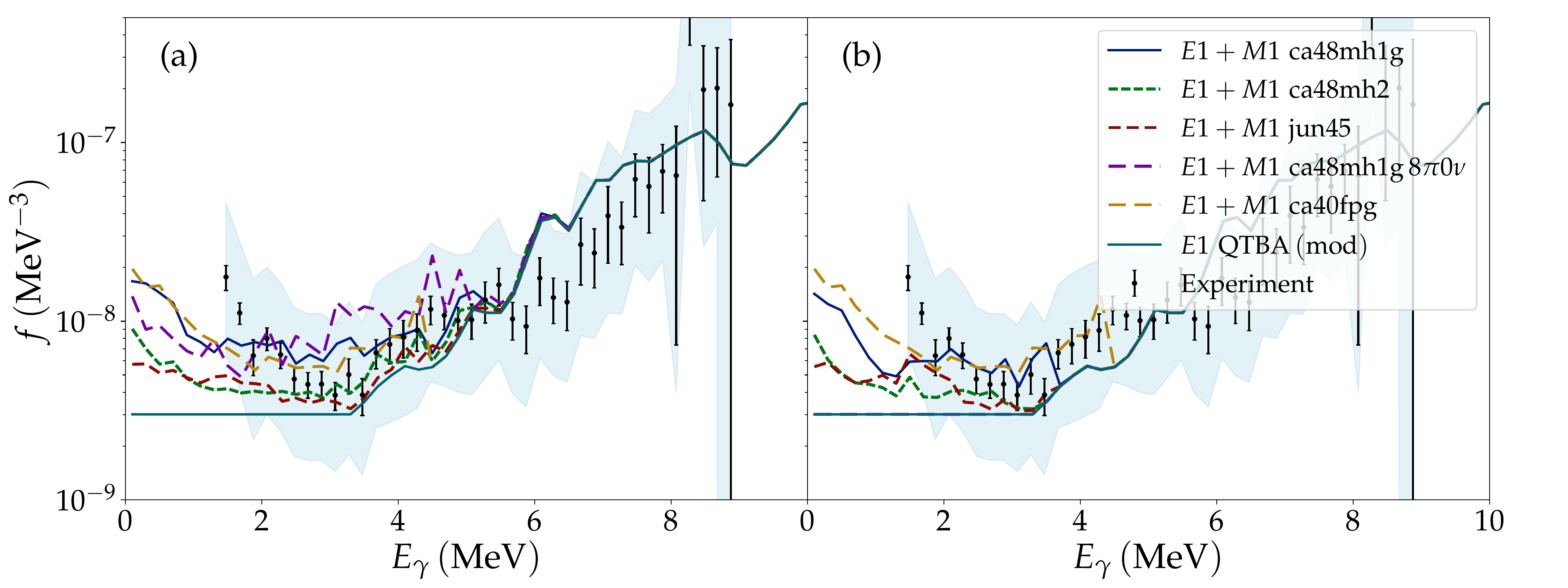}
\caption{(Color online) Sketch of a summed $\gamma$SF, $f_\mathrm{tot}(E_\gamma) = f_{E1}^\mathrm{QTBA(mod)}(E_\gamma) + f_{M1}^\mathrm{SM}(E_\gamma)$ for the different shell model calculations. The QTBA calculation has been replaced by a flat E1 strength for $E_\gamma < 3.5$ MeV, as suggested by recent shell model $E1$ calculations \cite{Sieja2017}.}
\label{fig:gSFsum}
\end{center}
\end{figure*}

Considering  transitions from initial states in the region $4.0 \leq E_x \leq 6.5$ MeV, we obtain the results shown in panel a) of Fig.~\ref{fig:gSF}.
In panel b) we show the $M1$ $\gamma$SF for transitions originating from $5^-$, $6^-$ or $7^-$ levels only, corresponding to the $^{70}$Co $\beta$ decay. The trend of the data for $1.5 \leq E_\gamma \leq 4.0$ MeV is well reproduced by all calculations, which all display an upbend peaking at $E_\gamma = 0$ MeV. The absolute value of the $M1$ strength is lower than the experimental total strength, although for {\scshape ca40fpg} and {\scshape ca48mh1g} it is within the error band. 
The QTBA calculations predict a drop of $E1$ strength towards $E_\gamma=0$. However, the QTBA is not a realistic model for the low-energy $E1$ strength because it is built on the ground state, so there are no low-$E_\gamma$ transitions available. A recent $E1$ shell model study for $^{44}$Sc predicts a flat behavior of the $\gamma$SF for low $E_\gamma$ \cite{Sieja2017}. If we assume a similar behavior for $^{70}$Ni, this brings the total strength into agreement with experiment, as sketched in Fig.~\ref{fig:gSFsum}.

There is also a discrepancy in the absolute strength between the different calculations, with {\scshape ca40fpg} and {\scshape ca48mh1g} having higher strength than {\scshape ca48mh2} and {\scshape jun45}. The highest strength functions seem to be in best agreement with the experimental total strength. 
The slope is however remarkably similar between all interactions, and consistent with an exponential function $Ae^{- E_\gamma / T}$. For {\scshape ca48mh1g} and {\scshape ca40fpg} we find $T \sim 1$ MeV, while $T$ is somewhat lower (higher) for {\scshape ca48mh2} ({\scshape jun45}), respectively.

Considering the $\beta$-decay selected $M1$ strength functions, the shape of the $\gamma$SFs do not change much, indicating that the selectivity of the $\beta$ decay does not introduce any significant bias in the experimental results. The reason for the abrupt drop in strength at $\sim 4$ MeV is the absence of direct transitions to low-lying states because of the $M1$ selection rules. 

We do see an upbend of equal slope even with the {\scshape jun45} interaction, {\it i.e.}~without any excited protons, in contrast to previous findings \cite{schwengner2013}. To investigate this further, we ran the {\scshape ca48mh1g} calculations with a different truncation, locking all neutrons and allowing all protons to excite (labeled $8\pi 0\nu$ in Fig.~\ref{fig:gSF}). 
The upbend is present here as well, with approximately the same slope, but disappears when applying the $\beta$-decay spin selection, because the protons-only truncation does not allow for any negative-parity states below $S_n$. 
\begin{figure*}
  \includegraphics[width=1.9\columnwidth]{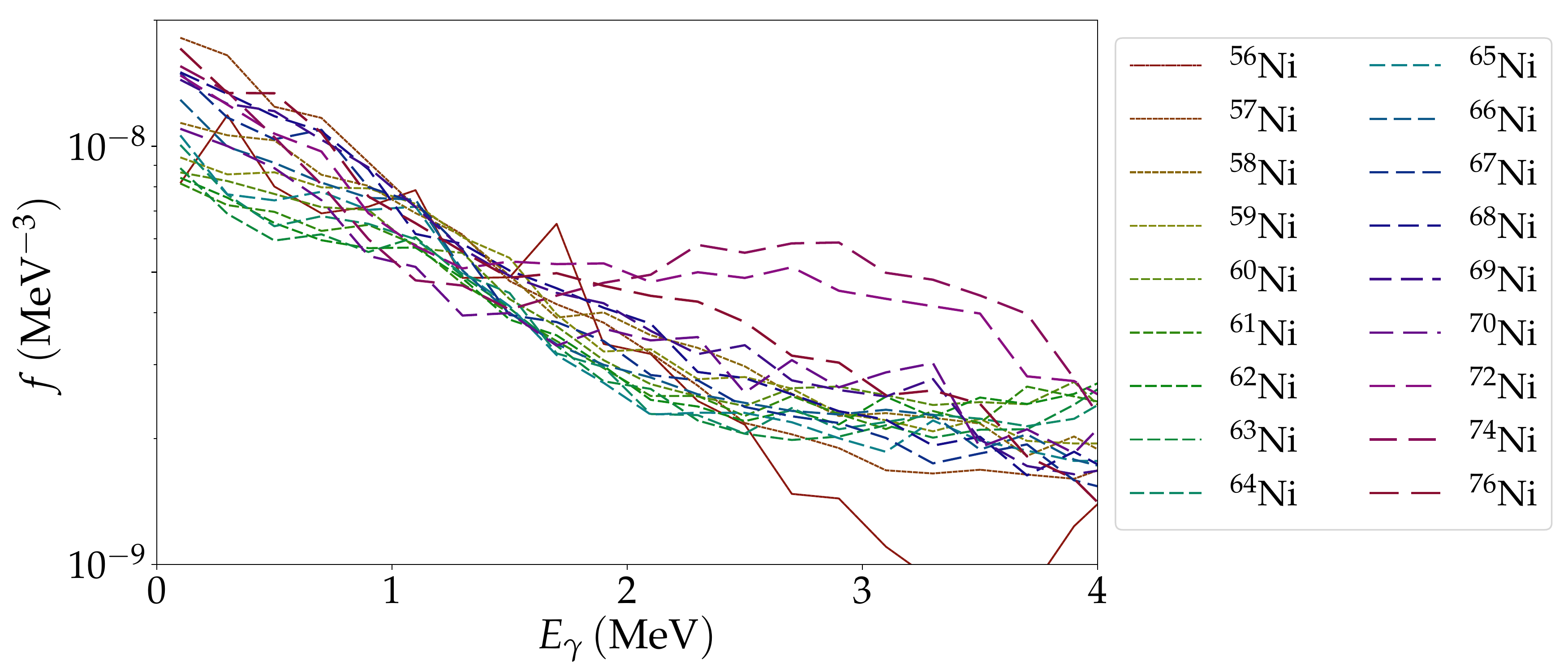}
\caption{(Color online) Calculated $M1$ $\gamma$ strengths for a wide range of Ni isotopes with the shell model, using the {\scshape ca48mh1g} interaction.	}
\label{fig:Nirange_gSF}
\end{figure*}
We find that the upbend is a remarkably robust feature for $^{70}$Ni within the shell model. 
During this work, we have explored a large parameter space of shell model interactions, of which only a subset is shown here. Try as we might, we did not find a single case where the upbend was not present. 

Finally, we expanded our theoretical scope and applied the {\scshape ca48mh1g} interaction, with the same proton truncation, to the whole isotopic chain, calculating $\gamma$ strength functions for $^{56-69,72,74,76}$Ni. We find an almost identical low-energy behaviour of the $\gamma$ strength across all isotopes, as shown in Fig.~\ref{fig:Nirange_gSF}.
The calculations agree with the low-energy behaviour of the experimental $\gamma$ strength for $^{64,65}$Ni~\cite{crespo2016,crespo2017}, as well as $^{69}$Ni \cite{spyrou2017}, the other neutron-rich isotope measured so far with the $\beta$-Oslo method. 
This is an intriguing indication that high intensities for low-energy $M1$ transitions could be a general feature of nuclei at high excitation energy, and could be expected all over the nuclear chart.

\section{Summary and outlook}
In this work, we have presented level-density and $\gamma$SF data on $^{70}$Ni extracted with the $\beta$-Oslo method. 
The experimental level density is found to be fully compatible with shell-model calculations including both positive and negative parity levels.

Our $\gamma$SF data  are  well reproduced by QTBA calculations for transition energies above $\approx 5$ MeV. 
On the other hand, at low transition energies, we find that the $\gamma$SF displays an upbend; thus $^{70}$Ni is the most neutron-rich isotope measured so far showing this feature. 
The upbend is described within the SM framework as an $M1$ component in the $\gamma$SF.  
SM calculations of the $M1$ $\gamma$SF are also performed for $^{56-69,70,72,74,76}$Ni. The results indicate that the upbend is a general trend for nucleon excitations in the quasi-continuum. 
Theoretical calculations for an even broader range of model spaces and heavier nuclei are ongoing and new computational methods are in development to investigate this further.

The experimental data will be available at \url{http://www.ocl.uio.no/compilation}.

\begin{acknowledgments}
A.C.L.~gratefully acknowledges funding of this research from the 
European Research Council, ERC-STG-2014 Grant Agreement No. 637686. 
A.~C.~L. acknowledges support from the “ChETEC” COST Action (CA16117), supported by COST (European Cooperation in Science and Technology).
Financial support from the Research Council of Norway, Project Grant No. 210007 (L.C.C., T.R., and S.S.) is gratefully acknowledged.
We would like to thank N.~Shimizu and M.~Hjorth-Jensen for valuable input to the shell model calculations, and T.~Engeland and E.~Osnes for stimulating discussions. 
A.C.L.~and J.E.M.~are grateful to the National Superconducting Cyclotron Laboratory at Michigan State University and to Lawrence Berkeley National Laboratory for their hospitality during parts of this work, as well as to the US-Norway Fulbright Foundation for Educational Exchange for supporting our stay in the United States.
{\scshape KSHELL} calculations were performed on the Abel, Stallo and Fram HPC clusters at the University of Oslo and the University of Troms{\o}, respectively, supported by the Norwegian Research Council. 
The work of O. Achakovskiy and S. Kamerdzhiev has been supported by the Russian Science Foundation, grant No.~16-12-10155.
D.L.B. acknowledges the support of LLNL under Contract No. DE-AC52- 07NA27344.
This material is based upon work supported by the Department of Energy/National Nuclear Security Administration under Award Numbers DE-NA-0003180, DE-NA0003221, and DE-NA-0000979.
This work was supported by the National Science Foundation under Grants No. PHY 1102511 (NSCL), PHY 1430152 (Joint Institute for Nuclear Astrophysics), PHY 1350234 (CAREER) and PHY 1404442. 
The LANL work was carried out under the auspices of the NNSA of the U.S. Department of Energy at Los Alamos National Laboratory under Contract No. DE-AC52-06NA25396. 

\end{acknowledgments}

\end{document}